\documentclass[aps,pra, twocolumn]{revtex4-1}
\usepackage{amsmath}
\usepackage{graphicx}
\usepackage{float}

\usepackage{graphicx}
\usepackage{dcolumn}
\usepackage{bm}
\usepackage{amsmath}
\usepackage{amssymb}
\usepackage[section]{placeins}
\usepackage{xcolor}
\usepackage{tikz}

\newif\ifredlined

\redlinedfalse
\ifredlined

\newcommand\circled[1]{\tikz[baseline=(char.base)]{
    \node[shape=circle,draw,inner sep=2pt] (char) {#1};}}
\newcommand\revcolor[1]{\color{#1}}
\else
\newcommand\circled[1]{}
\newcommand\revcolor[1]{}
\fi

\newcommand{\ax}{\alpha (x + \beta)}

\newcommand{\CP}[1]{Ci^{+}[#1]}

\newcommand{\pP}[1]{\CP{\ax}}

\newcommand{\be}{\begin{equation}}
\newcommand{\ee}{\end{equation}}
\newcommand{\ba}{\begin{align}}
\newcommand{\ea}{\end{align}}

\begin{document}

\title{
On the quantum tunneling time: Instantaneous, finite or probabilistic?
}

\author{S. Yusofsani, M. Kolesik}

\affiliation{
College of Optical Sciences, University of Arizona, Tucson, AZ 85721
}

\begin{abstract}
  Quantum particles interacting with potential barriers are ubiquitous in physics,
  and the question of how much time they spend inside  classically
  forbidden regions has attracted interest for many decades. Recent developments of new
  experimental techniques revived the issue and ignited a debate with often contradictory
  results. This motivates the present study of an exactly solvable model for quantum tunneling
  induced by a strong field. We show that the tunneling dynamics can depart significantly
  from the scenario in which the barrier-traversal time is zero or very small. However, our
  findings do not support the idea of a well-defined tunneling time either. 
  Our numerically exact results should help in finding a consensus about this
  fundamental problem.
\end{abstract}

\maketitle

\section{Introduction}

The tunnel effect~\cite{Hist,Mandelstam1928,Gamow} has captured the imagination of physicists from the early days of quantum mechanics.
Perhaps because of the lack of a classical analogue, one question in particular attracted a lot of attention;
namely how long is the time a quantum particle spends inside a potential barrier (see e.g.~\cite{MacColl32,Smith60,Buttiker82}),
i.e. in the spatial region that is inaccessible to it in classical physics. 
This problem was studied numerous times~\cite{Landauer94} in various physical contexts from the scattering on a one-dimensional
potential barrier~\cite{Buttiker83} to the mapping of electron trajectories in attosecond angular streaking experiments~\cite{KellerStreak,Landsman16}.

It is fair to say that the physics community has not yet completely agreed on the answer. On the contrary,
recent development of ultra-precise techniques in the strong-field physics~\cite{KellerStreak,Keller13}
reignited the debate as experiments produced often contradictory results.

The literature dealing with the tunneling dynamics is extensive. To give a few representative examples,
there are, very roughly speaking, three schools of thought concerning the subject. Some experiments produced
evidence that the time needed to traverse a potential barrier is zero,
negligible~\cite{Zero2019,backpropzero,Ni18,Keller13} or at least very small~\cite{KellerHelium}.
Others maintain that it takes a certain finite amount of time before a particle emerges from its ``quantum tunnel
through a potential barrier''~\cite{MPI-17,Landsman14,Landsman16,Serebryannikov:16,Keitel18,Yuan19}.
Yet other authors hold the idea that characterization in terms of a sharply defined
tunneling time is not suitable or useful for what is an inherently non-classical effect~\cite{Yamada99,Sokolovski18}.

One  possible reason that this controversy is difficult to settle is that sophisticated strong
field experiments~\cite{KellerStreak,Keller13} require an interpretation model~\cite{Smirnova15}
to give a meaning to the measured data,
e.g. one has to map the location of a detected free electron to its classical trajectory~\cite{backpropzero,Ni18}
in order to deduce the time
at which the electron was released from an atom. Theoretical approaches, such as numerical solutions to
the time-dependent Schr\"odinger equation~\cite{numerical,Lein18}, have their own challenges, too. Subtle differences between
definitions, including traversal, dwell, and reflection time~\cite{Buttiker83}, as well as the point of exit~\cite{Ni18,exitpoint},
and multielectron effects~\cite{Keller12} add further dimensions to the discussion.
\marginpar{\revcolor{red}\circled{III-1}}
          {\revcolor{red}
Some of the frequently discussed
approaches for the tunneling time are Larmur time, Buttiker-Landauer time~\cite{Buttiker82}, the Eisenbud-Wigner time~\cite{Wigner55}, Pollak-Miller time~\cite{Pollak84}, Wigner
time~\cite{Sokolivski07} and Bohmian time~\cite{Kocsis11,Wiseman07}. While the first four of these do not have a straightforward application in the situation discussed
in this work (due to differences in the physical setting and/or simplicity of our model), 
Wigner and Bohmian times could apply to our scenario, with the Wigner time being perhaps the most relevant.
            }

Inspired by the ongoing debate, we present a theoretical study of a simple, but exactly
solvable model that allows one to accurately characterize the time-dependent wavefunction of a tunneling particle.
\marginpar{\revcolor{red}\circled{III-2}}
\marginpar{\revcolor{red}\circled{II-3}}
          {\revcolor{red}
            Of course, simple systems have been employed in this problem before (see e.g.~\cite{Klaiber20},
            and the scenario investigated in ~\cite{onedmodel} is
            similar in spirit to ours). Here we want to concentrate on tunneling invoked by an external
            field.
}
Our approach makes it possible to study the dynamics in the regime of a nearly opaque, spatially very long barrier
that is difficult to address by other methods, and which greatly emphasizes the quantum nature of the effect.
In particular, we are able to investigate the dynamics for very weak field strengths, where the distinction
between instantaneous and delayed tunneling is clearer, and where it is easier to identify the effective
exit point from the quantum tunnel.

\section{Model} 

The rationale for choosing the model for our investigation is
the need to eliminate the uncertainties that one inevitably encounters in
the numerical solution and in the interpretations of results for a more realistic system.
We consider a toy model that can be solved exactly, and the solution of which can
be accurately numerically evaluated. Let us assume a Stark problem given by a Hamiltonian
for a one-dimensional particle,
\begin{align}
  H &= -\frac{1}{2} \frac{d^2}{d x^2}  - V_0 \ \ \ \  \text{for} \ \ \ -L \le x \le 0  \nonumber \\
  H &= -\frac{1}{2} \frac{d^2}{d x^2}  - F x \ \ \ \  \text{for} \ \ \  x > 0 \ ,
\label{eqn:hamiltonian}
\end{align}
where the field strength $F>0$ and the depth of the potential well in the left half-space is constant
(with positive $V_0$). To select the domain of Hamiltonian (\ref{eqn:hamiltonian}),
we assume $\partial_x \psi(x\!\!=\!\!-L)\!=\!0$.
At $x\!=0\!$ we require the functions belonging to the domain of the operator to be continuous,
together with their first derivative.

This model is inspired by a metal nano-tip exposed to a strong external field as realized in experiments
by irradiation by strong optical pulses~\cite{nt_ropers,nt_twocolor}. Here we consider
a constant field strength, and concentrate on time-scales much shorter than the optical cycle.
The potential well represents a partially filled conduction band, so that the energy
of $W = -V_0$ represents the bottom of the band, and the states in the vicinity of the Fermi level will
contribute most to the tunneling current. While the limit $L\to\infty$ can be taken, we shall evaluate
our illustrations for a finite $L$, and note that a concrete choice of its value plays no significant
role in our observations. 

We will examine this model also in a complementary limit representing a different physical situation.
Namely, we take a small-$L$ and large $V_0$ limit, such that there exist exactly one bound state with
the energy equal to $1/2$. This is to mimic quantum tunneling from atoms exposed to strong fields.

The scenario we are to examine is as follows. We assume that the system is prepared in the absence of the
external field, i.e. we have $F=0$, and the initial wavefunction is selected as one of the bound eigenstates.
We denote its energy by $Q$.

\begin{figure}[H]
  \begin{center}
  \includegraphics[width=0.48\textwidth,clip]{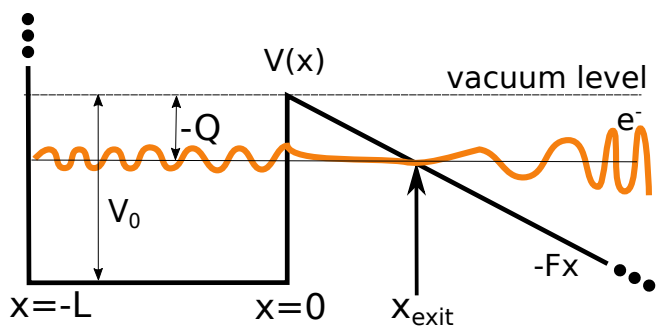}    
  \caption{Color online.
    Potential function for a 1D model of a metal nano-tip.
    For the initial condition given by a zero-field stationary state with energy of $-\vert Q\vert$,
    we aim to calculate the time-dependent wavefunction of the tunneling particle for locations
    beyond the classical tunnel exit at $x_\text{exit}$.
    }
  \label{fig:tipmodel}
  \end{center}
\end{figure}

Then, at time $t=0$ we suddenly switch the field on to $F>0$, and then follow the evolution of the
wavefunction. Any positive value of the the field strength $F$ creates a finite but possibly broad potential barrier
through which the initial state can tunnel toward $x\to\infty$. We are particularly interested in how long
it takes for the particle to appear at the classical exit from the tunnel, i.e. at the location $x_\text{exit}=-Q/F$.
More generally, it is desirable to understand the dynamics of the wavefunction in the classically allowed region
$\psi(x>x_\text{exit},t>0)$. For example, from the application standpoint it is important to understand the emitted
electron bunches, including any limits for the  duration of such pulses.
The situation is schematically depicted in Fig.~\ref{fig:tipmodel}.

\noindent
{\bf Initial wavefunction.} \\
The initial condition is a zero-field eigenfunction corresponding to a negative energy
$Q\in (-V_0,0)$. Convenient parameterizations, utilizing a real-valued wavenumber $k_Q = \sqrt{2 (V_0 + Q)}$,
can be written as
\begin{align}
\psi_Q(x) &=  \frac{\cos( k_Q (L + x) )}{\cos(k_Q L)} \ \ \text{for} \ \ x<0 \ \ , \ \ \ \ \\ \nonumber
\psi_Q(x) &=  \exp( -\sqrt{-2 Q} x) \ \ \text{for} \ \ x>0 \ ,
\end{align}
and the eigenvalue equation for energy $Q$ is
\be
k_Q \tan( k_Q L ) =  \sqrt{-2 Q} \ .
\ee
The evolution of the system starts from such a bound state after a sudden switch of the
field $F$ from zero to a finite positive value. The instantaneous switch-on of the field gives a clear
meaning to the time when the evolution starts.
If there was a finite ramp-up time of $F(t)$, it would not be clear exactly at what time does the
tunneling process commence. Despite the fact that the tunneling probability grows exponentially with the field,
the ramp inevitably introduces an uncertainty into the interpretation of $t=0$.
On the other hand, for an instantaneous turn on of the field, the dynamics we are going to encounter is
non-adiabatic as we will appreciate shortly.


\noindent{\bf Spectrum and eigenfunctions in non-zero field.}\\
Needless to say, ours is a text-book, exactly solvable system for which all Hamiltonian eigenstates can be easily
obtained.
For any positive $F$, however small, the spectrum of our system becomes continuous, and it encompasses the whole
real axis. The following is a suitable parameterization of the energy-eigenfunctions we shall utilize in what follows:
\begin{align}
  H \psi_W(x) = &W \psi_W(x) \ , \nonumber \\
  \psi_W(x) = &\frac{\cos(k_W (L+x))}{\cos(k_W L)} \frac{1}{N \sqrt{D^+ D^-}} \ \ \  \text{for} \ \ \  x < 0 \ \ \ \nonumber \\
  \psi_W(x) = &\frac{-i}{N}\sqrt{\frac{D^-}{D^+}}  Ci^+(\alpha (x + W/F)) + \\ \nonumber
  &\frac{+i}{N}\sqrt{\frac{D^+}{D^-}} Ci^-(\alpha (x + W/F))
  \ \ \ \text{for} \ \ \ x > 0  \ , 
\end{align}
where $k_W = \sqrt{2 (V_0 + W)}$, $\alpha = -(2 F)^{1/3}$ and $N=2^{7/6} F^{1/6}$ is a normalization factor. Functions $Ci^\pm$ are combinations of the
Airy functions,
\be
 Ci^\pm(z) = Bi(z) \pm i Ai(z) \,
\ee
and are chosen as such in order to express $\psi_W$ as a superposition of the incoming and outgoing waves.
Specifically, $Ci^+(\alpha (x + W/F))$ behaves as an outgoing wave in the region of large positive $x$.

Functions $D^\pm(W)$ are determined from the requirement of smoothness at $x=0$. Asking for the continuity of
the wavefunction and of its first derivative at $x=0$, one obtains
\begin{widetext}
\begin{align}
  D^+(W) &= \frac{\pi}{2 } \left( Ci^{+'}(\alpha W/F) - \frac{k_W}{\alpha} \tan(k_W L)  Ci^{+}(\alpha W/F) \right) \nonumber \\
  D^-(W) &= \frac{\pi}{2 } \left( Ci^{-'}(\alpha W/F) - \frac{k_W}{\alpha} \tan(k_W L)  Ci^{-}(\alpha W/F) \right) \ .
\end{align}
\end{widetext}
  Note that the zeros of these expressions, when analytically continued to complex plane, determine the location of
  the Stark resonances for the model under consideration.
When $W=z$ is chosen such that $D^+(z)=0$, the incoming part of the wavefunction vanishes, while the outgoing one has poles
at a complex-valued energies. We will utilize the outgoing resonances to construct part of the wavefunction of the 
particle as it tunnels through the potential barrier.

Note that the ``atom-model'' limit of $L\to 0$ and $V_0\to\infty$ can be taken in the above formulas by taking
\be
k_W \tan(k_W L) \to 1 .
\label{eqn:delta-limit}
\ee
This limit introduces a contact or Dirac-delta interaction at the origin, and fixes the energy of the single bound state to $-1/2$.

\bigskip

\noindent{\bf Expansion in energy eigenstates}\\
We start by formulation of the time-dependent solution in the standard way, using the completeness of the
Hamiltonian eigenfunctions.
The eigenstates $\psi_W(x)$ can be normalized to the Dirac-delta function in energy $W$,  so that we have
a unity decomposition guaranteed to exist for the self-adjoint operator,
\be
\int\!  \psi_W(x) \psi_W(y) dW = \delta(x-y) \ ,
\ee
which is the completeness relation that allows one to express an 
arbitrary initial wavefunction $\phi(x)$, like so
\be
\phi(x)\! =\!\! \int\!\! \delta(x\!-\!y) \phi(y) dy =\!\!  \int\!\! \psi_W(x)\!\! \int\!\! \psi_W(y) \phi(y) dy dW  ,
\ee
     where we used the fact that $\psi_W$ is real.
The evolution of this initial condition at later times can be described with the expansion
written like this
\be
\psi(x,t)\!  =\!\!  \int\!\! \frac{e^{-i t W} \psi_W(x)}{\sqrt{D^-(W) D^+(W)}} A(W) dW
\ee
with the overlap integral
\be
A(W) =  \int\!\! \sqrt{D^-(W) D^+(W)} \psi_W(y) \phi(y) dy .
\ee
The $D^\pm$ factors were distributed in the above such that we have analytic functions
when continued into complex plane.

Interested in the tunneling part of the solution
for $x>0$, we can write the outside component  as 
\be
\psi\! =\!\!
\int\!\! \frac{i e^{-i t W}}{N}\!\!  \left[\! \frac{  Ci^-(\alpha x\!+\! \frac{\alpha W}{F}) }{D^-(W)} - \frac{ Ci^+(\alpha x\! +\!\frac{\alpha W}{F}) }{D^+(W)}\! \right]\!\!  A(W) dW.
\label{eq:psit}
\ee

The spectral amplitude $A(W)$ can be split into
$A(W) = A_\text{I}(W) + A_\text{O}(W)$ where the overlap integral between $\psi_W$ and $\phi$
consists of the ``inside'' and ``outside'' (of the well) contributions.
Specializing this for the chosen initial wavefunction $\psi_Q(x)$, we have an exact
\be
A_\text{I}\!=\! 
\frac{\sqrt{-2 Q} - k_W \tan(k_W L) }{2 N (Q-W)}
\ee
and
\begin{align}
\label{eqn:overlap}
  &A_\text{O}\!=\! \frac{-i}{N} \int_{0}^\infty\!\!\!\! dy\  \exp[-\sqrt{-2 Q} y] * \\ \nonumber
&\left[ D^-(W) Ci^+\left(\alpha (y + \frac{W}{F})\right) - D^+(W) Ci^-\left(\alpha (y + \frac{W}{F})\right) \right] 
\end{align}
that must be calculated numerically.

\bigskip
\noindent{\bf Expansion in resonant states}\\
The expansion of the time-dependent wavefunction in terms of energy eigenstates
(\ref{eq:psit})
can be in principle evaluated. Unfortunately, numerical calculation of the integral in (\ref{eq:psit})
is next to impossible due to extremely fast variation
of the integrand. This is a challenge especially for very weak field $F$ that pulls the arguments
of the embedded Airy functions to infinity.  The fast integrand variation is caused by poles 
located extremely close to the real axis. These poles correspond to the
Stark resonances, which may be viewed as eigenstates of a non-Hermitian Hamiltonian which
has the same differential expression as (\ref{eqn:hamiltonian})
but has its domain specified by the asymptotic boundary condition which requires that the functions that
belong to the domain behave as outgoing waves (also known as the Siegert boundary condition).

We  proceed to evaluate the formally exact expression for $\psi(x,t)$, 
by deforming the integration contour from that following the real axis to one for which
the difficult to calculate part of the integral can be obtained from the Stark poles.

We choose an integration contour $C\!=\!\{z\in\mathbb{C};\text{Im}\{z\}\!=\!-s\}$ that
parallels the real axis in the lower complex half-plane. Utilizing the residue theorem,
the expression for the outside wavefunction can be written like so
\begin{align}
\psi(x,t)\! =\!   -2 \pi  \sum_{p}   e^{-i t W_p}
\frac{Ci^+(\alpha (x + W_p/F))}{N D^{+'}(W_p)} A(W_p)\nonumber \\
+
\int_C\!\! \frac{i e^{-i t z}}{N}\!\!  \left[\! \frac{  Ci^-(\alpha x\!+\! \frac{\alpha z}{F}) }{D^-(z)} - \frac{ Ci^+(\alpha x\! +\!\frac{\alpha z}{F}) }{N D^+(z)}\! \right]\!\!  A(z) dz
 \label{eq:series}
\end{align}
Here the set of poles $W_p$ that were crossed by deformation of the contour is finite and it depends on the
choice of the parameter $s$. The discrete sum above is a resonant-state expansion, and its purpose
here is to replace the part of the integral that is most difficult to evaluate.

Resonant series expansions similar to the discrete part of (\ref{eq:series}) were successfully
used in systems exhibiting spontaneous decay without the external field (e.g.~\cite{garcia-calderon}).
The situation is somewhat different for the Stark resonant states that arise due to
the homogeneous external field. We therefore think it may be illustrative to discuss the general structure of
the  poles that are relevant for our resonant-state series.

\begin{figure}[t]
  \begin{center}
    \includegraphics[width=0.40\textwidth,clip]{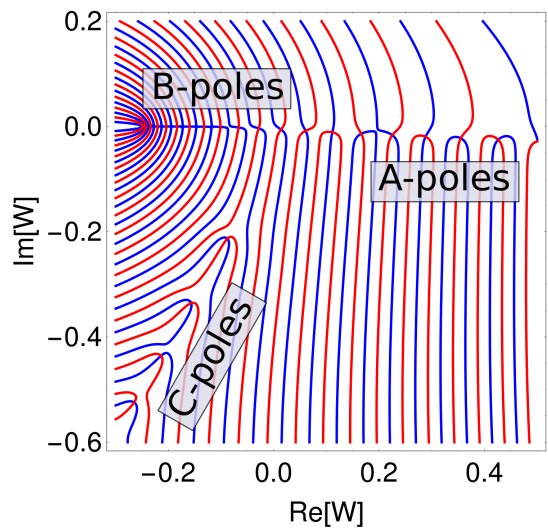}
  \caption{Color online.
    Solutions to the resonant-state eigenvalue equations. The contours
    are zero lines for the real and imaginary parts of $D^+(W)$, and their intersections
    indicate the location of the resonant poles $W_p$. Three different families of
    solutions are indicated here and discussed in the text.
    }
  \label{fig:pole-map}
  \end{center}
\end{figure}

There are three
families of poles distributed in the lower half of the complex plane (with their counterparts related to the incoming
resonant states living in the upper half-plane). They are illustrated in Fig.~\ref{fig:pole-map}.
The figure shows the contours defined by Re$\{D^+(W)\}=0$ in blue and Im$\{D^+(W)\}=0$ in red.
Thus, the intersections
of these contours mark the locations of the resonant-state poles. The first set are the
B-poles; they correspond to the metastable states that arise from the zero-field bound states. The imaginary
parts of these complex energies are tiny and therefore invisible on the scale of this figure. The second set of poles,
marked with $A$ in the figure, correspond to the resonant states  similar to the positive-energy states at $F=0$.
There are infinitely many of these poles, located below the real axis, with the imaginary part growing for the resonances
with larger real parts. The third set of poles, indicated by $C$ in the picture, are located along the Stokes line of
Airy functions. Again, there is infinitely many of them, each in the vicinity of zeros of $Ci^+(\alpha z/F)$.

It should be noted that the parameters for this illustration were chosen such that only a few poles appear
in each family. Specifically, this requires an extremely strong field,
and a small potential well that only accommodates a few bound states.

We choose our contour $C$ to run  close to the real axis, but ``below'' B-poles. As $C$ continues into the 
half-plane $\text{Re}\{z\}>0$, it eventually crosses the line of A-poles, so the resonant-state expansion part
of (\ref{eq:series}) is limited to those poles that are located between the real axis and contour $C$.
The specific choice of $C$ is informed by the numerics involved. As the contour drops deeper into the lower plane,
the influence of the B-poles becomes less severe, and their contribution is replaced by the resonant-state sum.
However, Airy functions grow exponentially fast away from the real axis, and this means that there exist
huge cancellations in the numerical evaluation of various terms contributing to the contour integral.
It is therefore most practical to keep $C$ not too distant from the real axis. Results shown here were obtained with
s=3/100 which required about sixty poles contributing to the resonant-state series. We have performed all our calculations
for several different values of $s$ in order to verify that the results are indeed independent of the choice for the contour.

\section{Results}

\subsection{Tunneling from a discrete state}

We start our illustrations with the ``atom model'' case since it is much less computationally demanding.
For this we take the large-$V_0$ plus small-$L$ limit specified in (\ref{eqn:delta-limit}), so that there remains a single bound state
with the energy of $Q=-1/2$ and the wavefunction $\phi(x)=e^{-x}$, that serves as our initial condition.

\begin{figure}[h!]
 \begin{center}
       \includegraphics[width=0.40\textwidth,clip]{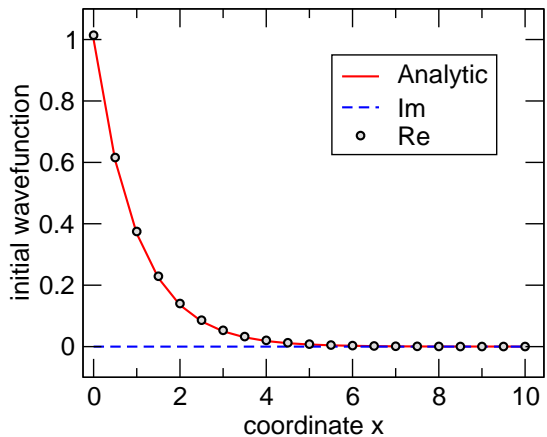}
    \caption{Color online.
      The initial wavefunction of the particle as reproduced by the mixed representation
      utilizing a contour integral and a subset of outgoing resonant states~(\ref{eq:series}).
    }
  \label{fig:initial-phi}
  \end{center}
\end{figure}

In order to check that our expansion and numerical integration in (\ref{eq:series}) are properly
normalized and can be evaluated with sufficient accuracy, we verify that for $t=0$ we indeed recover
the initial wavefunction. This is illustrated in Fig.(\ref{fig:initial-phi}). The imaginary part
of the recovered wavefunction should vanish, and our calculations gives values of the order of $10^{-8}$.
The error in the real part is on the level of a percent at the origin where it is largest.

Concentrating on the question of the tunneling time, we observe the probability density as it evolves at a
fixed point of observation. If one assumes that the so-called simple-man's picture of quantum tunneling applies,
then we expect that the particle appears at $x_\text{exit}=-Q/F$ instantaneously and  it has zero velocity
at the exit from the tunnel. Then, subject to acceleration by the external field, it moves away and arrives at
a chosen observation location $x_\text{obs}$.  The expected time of arrival based on this scenario
is $t_\text{eta}=\sqrt{2(x_\text{obs}-x_\text{exit})/F}$.

\begin{figure}[h!]
 \begin{center}
       \includegraphics[width=0.40\textwidth,clip]{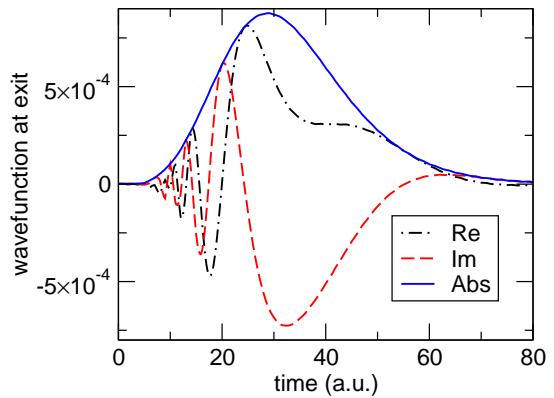}
    \caption{Color online.
      Tunneling from a single bound state; particle wavefunction is shown versus time at the location
      corresponding to the classical tunnel-exit point. 
    }
  \label{fig:wf-at-exit}
  \end{center}
\end{figure}

First we evaluate the full wavefunction at the position $x=25$ which corresponds
to the classical exit from the tunnel for the given applied field $F=1/50$.
Figure~\ref{fig:wf-at-exit} shows that the tunneling particle appears at the
classical exit not instantaneously, but delayed. The real and imaginary parts
of the wavefunction evaluated versus the observation time indicate that at earlier times
the particle arrives most likely with a higher velocity that at later times.
\marginpar{\revcolor{red}\circled{II-3}}
{\revcolor{red}
  This is in line with previous works, e.g. ~\cite{Xu18}, that have also shown that for a 1D model the tunneling ionization  produces a non-zero outgoing momentum at the tunnel exit.
}
Clearly, the simple-man's scenario does not apply for this specific situation as there is a pronounced delay
and the arrival time exhibits a broad distribution.


\begin{figure}[h!]
 \begin{center}
       \includegraphics[width=0.40\textwidth,clip]{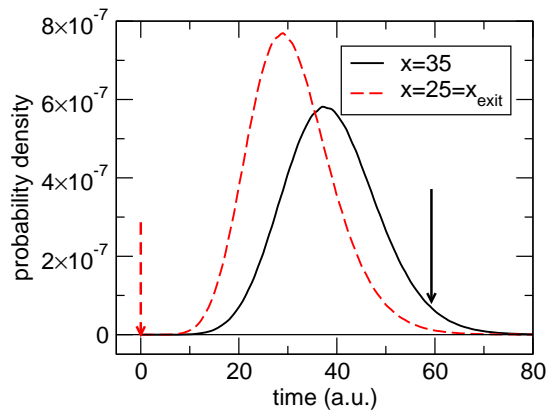}
  \caption{Color online.
    Tunneling particle probability density versus time for two different observation locations $x$
    for a model with a single bound state taken as the initial wavefunction. The dashed line
    corresponds to the location at the classical tunnel exit.
    Arrows mark the expected time of arrival for a particle that obeys the simple-man's scenario.
  }
  \label{fig:density-vs-time}
  \end{center}
\end{figure}

In order to appreciate the arrival-timing of the particle at different observation locations,
Fig.~\ref{fig:density-vs-time} depicts the observed probability density versus time
as detected at two locations. Arrows mark the expected time of arrival at the given location
under the assumption that the simple man's scenario holds. When the peak of the probability density
pulse is adopted as the location of the classical trajectory, 
it becomes evident that
the particle arrives early at more distant locations, indicating that the initial velocity
at the classical exit does not vanish, which is also in contrast to the simple man's scenario.

\subsection{Tunneling from a quasi-continuum of states}

To illustrate our observations concerning the dynamics of the tunneling particle
for the model of a metal nano-tip, we choose
for $L=100$, $V_0=25/68$, $Q=-0.1848\ldots$ and a very weak field $F=1/100$. The depth of the
potential well $V_0$ and the energy $Q$ of the initial stationary state are motivated by metal
nano-tips, with $Q$ taken roughly corresponding to the typical work function of common metals.
In contrast, the above field value (in atomic units) is significantly smaller than the typical
fields achieved by irradiation by femtosecond optical pulses. As alluded to in the introduction,
we concentrate on weak fields in order to make the potential barrier nearly opaque, and thus ``amplify''
the potential deviations from the scenario of the instantaneous tunneling.

The first step in numerical evaluation of the expansion in (\ref{eq:series}) is to obtain the locations of poles,
or solutions to the equation $D^+(W) = 0$. This can be done by initiating the search around the energies close to the bottom
of the potential well. Having found two and more poles one can estimate the location of the next by extrapolation,
and subsequently find the root with working precision of up to two hundred digits. The high numerical accuracy
during this and subsequent calculations is a must in order to obtain converged results. Depending on the contour
chosen,  several tens of resonant poles $W_p$ are needed for the chosen size of the potential well.


The next step consists in calculating the different terms that originate in the integrand of (\ref{eqn:overlap}),
in particular the wavefunction overlaps $A(W_p)$ and the pole-residue values $D^{+'}(W_p)$. Obviously, high-precision
arithmetic must be employed.
%
%
Having found the complex resonant energies and the corresponding wavefunction overlaps together with the
pole residues, we can evaluate the resonant-expansion part of the wavefunction.

Evaluation of the contour integral is straightforward but it requires a lot of numerical effort.
We found that accurate tabulation of the integrand along the contour with very fine discretization along $z$
helps to speed up the calculation for the wavefunction as a function of time.

\begin{figure}[t]
  \begin{center}
        \includegraphics[width=0.40\textwidth,clip]{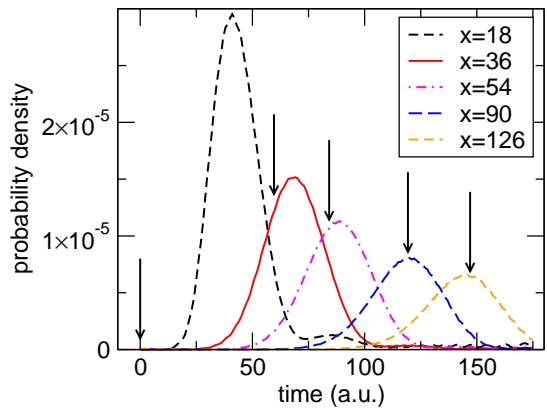}
  \caption{Color online.
    Tunneling particle probability density versus time for different observation locations $x$.
    Arrows mark the expected time of arrival for a particle that obeys the simple-man's scenario.
    }
  \label{fig:pulses}
  \end{center}
\end{figure}

Figure \ref{fig:pulses} shows selected results for one of our simulations.
For the chosen parameters, the exit from the tunnel is at $x_\text{exit}=18$,
and the full black curve corresponds to this observation location. It reveals that the maximum of the probability
density arrives significantly delayed with respect to $t_\text{eta}=0$. Obviously, in this case the tunneling
time seems to be finite, and in fact rather far from instantaneous.

Several other curves in Fig.\ref{fig:pulses} show similar results for the observation
locations at increasing distances from the tunnel exit. The arrows in the plot indicate the
expected time of arrival in each case. One can see that while the particle seems to be delayed
when observed close to the tunnel exit, it arrives earlier than expected when detected far from
the tunnel. In other words, it moves faster than expected and this is due to the fact that
in contrary to the simple-man's model 
it exhibits a non-zero velocity at the exit.
%
One could argue that the classical exit point $x_\text{exit}$ must not be taken as given by the
nominal value $-Q/F$, but should be adjusted. Indeed, classical trajectory fitted to our numerical
results would indicate the effective tunnel-exit point at $x \sim 5$ and the escape velocity
of 0.12 in atomic units.

\begin{figure}[t]
  \begin{center}
    \includegraphics[width=0.4\textwidth,clip]{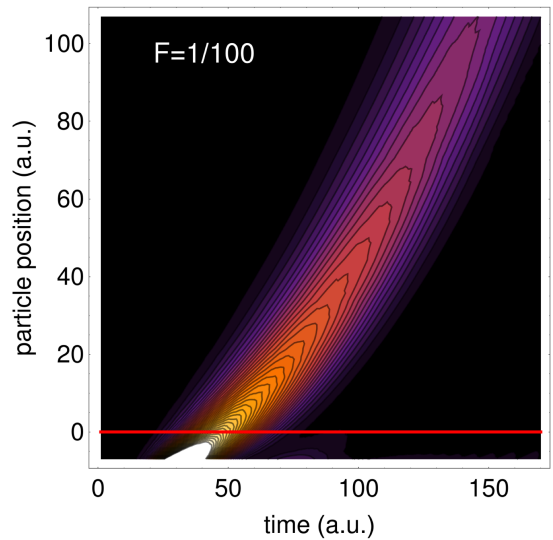}
  \caption{Color online.
    Probability density for the tunneling particle versus time and observation location.
    The red horizontal line indicates the location of the tunnel exit for the given external
    field strength.
    }
  \label{fig:pulse-map}
  \end{center}
\end{figure}

\marginpar{\revcolor{red}\circled{II-1,2}}
          {\revcolor{red}
A note concerning the interpretation of the curves shown in
Fig.~\ref{fig:pulses} might be in order. The figure shows that the amplitude of the pulse
decreases with the distance from the origin, while the duration only increases slowly.
It might therefore seem that  the total probability of finding the particle inside of the pulse
may not be conserved. However, one has to consider the fact that the particle accelerates in
the external field, and the pulse in fact delivers the same accumulated
probability independently of the distance. This can be verified accurately
when the wavefunction (or, more accurately, the probability density) is
integrated over space for a fixed time; In our illustration the probability
is conserved with an accuracy of a few parts in thousand.

However, one should also note that our calculation evaluates the wavefunction as
a whole and that is why it also contains a portion that remains bound
on the time-scales shown in our illustrations.
We think that the tiny variations
in the total probability transported by the pulse actually reflect the
finite accuracy achieved in our calculation, and that the temporal
variations of the "bound-part" of the wavefunction are small in comparison.
So keeping in mind the finite accuracy of the calculation, it is reasonable to
say that these pulses represent mostly particles that are escaping toward infinity.
}
  
In order to visualize the particle trajectory, Fig.~\ref{fig:pulse-map} shows the probability density
plotted versus time and observation location. One can see that the particle's probability density concentrates
along the parabola reflecting its classical acceleration. However, the slope at the intersection
with the red horizontal line which marks $x_\text{exit}$ indicates that the particle appears
here delayed and with a non-zero velocity.
Interestingly, the figure also brings to light that a
well-defined wave-packet forms already before it reaches the classical exit point from the tunnel. 
\marginpar{\revcolor{red}\circled{II-3}}
{\revcolor{red}{In effect, there seems to be a distribution of the tunnel exits~\cite{Ni18b} and corresponding velocities.}}

\marginpar{\revcolor{red}\circled{III-1}}
          {\revcolor{red}
      This is  similar to the Wigner tunneling picture~\cite{wigner}, but in our case a range of energies contribute to the
wavefunction giving rise to a ``fuzzy'' particle trajectory, with parameters potentially
different than those calculated for a fixed-energy propagator~\cite{wigner} suitable for an adiabatic regime.
Also note that ours is a non-adiabatic regime when mapping to a single classical trajectory may not be fully adequate~\cite{keitel}.
}

Our results also illustrate that the wavepacket exhibits a quite well-defined duration,
which can be perhaps surprising given that there is no inherent time-scale imposed by the
external field as there would be in the excitation by an optical pulse.
However, one can appreciate that there is a natural time-scale imposed by the energy spectrum
of the states that contribute to the tunneling current. The broader
is the initial energy spectrum the shorter is the resulting particle pulse. 

The duration of the emitted pulse is relevant for the applications of metal nano-tips as super-fast sources
of electron bunches~\cite{nt_twocolor,nt_control}. While ours is at best an idealized model for a fast electron source, it
is reasonable to look at this result as an estimate for the ``fastest achievable electron packet.''
Moreover, our illustration clearly shows that the resulting particle pulse consists of a spectrum of energies
and, accordingly, suffers from dispersion which will spread-out the pulse upon further propagation.

\section{Discussion}

This section is devoted to a brief discussion of a few technical issues relevant for the calculations underlying our results.

First of all, it should be emphasized that the above example represents a generic behavior that we have obtained
for a number of various model settings. We have found that whenever the field strength is
weak we see very pronounced deviations from the scenario of instantaneous tunneling.

Yet another aspect the reader might be wondering about is that the expected exponential decay of the
meta-stable state born out of the initial stationary state is not evident in our results.
The explanation has to do with the fact that we look at relatively weak field strengths; The evanescent tail of the
initial bound state is exceedingly small at $x=x_\text{exit}$ unless its energy $Q$ is close to the very top
of the potential well. The outgoing probability current induced to this state by the external field also occurs at
much slower time scale and it therefore appears negligible on the background of the fast wavepacket generated by the
suddenly imposed field. 
For the given conditions, the decay rate of the metastable states energetically close to the initial
wavefunction is of the order of $10^{-9}$. The ``tunneling pulses'' seen in our examples exhibit velocities
on the order of unity and result
in roughly $10^{-5}$ ionization rate, so their contribution dominates by about four orders of magnitude. 
However, it is possible to create situations in stronger fields where these slow and fast
components of the tunneled wavefunction are comparable.

Last but not least we touch upon the the non-adiabatic aspect of the tunneling dynamics seen in our calculations.
Clearly, it is the instantaneous switch-on of the field that is responsible for the broad energy spectrum of the
resulting wavefunction at time $t\to 0^+$. 
One could ask if a more adiabatic turning on of the field can result in a more sharply
defined tunneling time. Intuition suggests that when the field is turned on slowly the higher-energy components
of the resulting state will be reduced and, consequently, the wave-packet will broaden more in time due to
its smaller bandwidth. Moreover, the slow ramp-on
of $F(t)$ introduces significant arbitrariness into the meaning of $t=0$ because only a sudden switch-on
gives a clearly defined initial time for the evolution. So, while in terms of non-adiabadicity ours is a most extreme
scenario that can only be approximated in an experiment, it shows that the way the system is excited introduces yet
another aspect that complicates the notion of the classical barrier-traversal time.

\section{Conclusion}

We have presented an exactly solvable model for an electron tunneling from a
metal nano-tip exposed to an external field. Taking advantage of the non-Hermitian
reformulation of the time-dependent problem, we were able to calculate the 
wavefunction for the field strengths that present a difficult challenge for more standard methods,
including numerical solutions of the time-dependent Schr\"odinger equation.
The time-dependent wavefunction revealed that the energy spectrum of the system imposes
an inherent lower limit on the duration of the electron pulse emitted by the tip even
when the field turns on suddenly. We have also seen that the energy spectrum of the emitted particle
is wide and the tunneling wave-packets therefore experience strong temporal dispersion.

However, our main motivation for this study was the currently debated question of the tunneling time,
which some claim to be instantaneous, while others present evidence that it must be finite, and still others
maintain that the quantum nature of the tunneling problem precludes a meaningful description in  classical terms.
As for the notion of the tunneling time, our results show that 
for our particular examples in
one dimension and for a weak external field the tunneling dynamics
exhibits a pronounced delay between the sudden switch-on of the field and the time when
the particle can be detected at the point of the classical exit from the potential barrier.
Moreover, the so-called simple man's scenario of quantum tunneling also does not apply because the
particle has a significant velocity when it appears at the classical exit point. So in this exactly
solvable system, one can safely say that the tunneling time is not instantaneous.
However, our results do not support the idea of a well-defined tunneling time either.
There are at least two reasons for this. First, it is obvious that at the point of classical exit,
the temporal distribution of the probability density ``pulse'' is actually broader than the
apparent tunnel time as defined by the time of arrival of the peak.
In other words, the tunneling time
  is a stochastic quantity that perhaps could be better described by a probability distribution.
Second, we have seen that
a well-defined moving wave-packet forms actually {\em before} the particle reaches the
classically allowed region. In hindsight this is not very surprising given the energy uncertainty
caused by the fast switch-on of the field. Because of the energy spread, some components of the wavefunction
experience the potential barrier as a classically accessible region. This makes the very notion
of the tunnel exit rather ill-defined. By extension the utility of the tunneling time itself
is limited at best. However, the time-dependent wavefunction exhibits a well-defined although
fuzzy trajectory, and in this sense the behavior in our illustrations resembles the Wigner
scenario~\cite{wigner}.
\marginpar{\revcolor{red}\circled{III-1}}
          {\revcolor{red}
The question of how the results of this paper compare {\em quantitatively} to Wigner`s time is
interesting, and will be addressed elsewhere.
}

To conclude, we do not think that observations based on a simple 1D model can be an arbiter
in any way between the different schools of thought and experiments concerning the problem of
the tunneling time. We hope, however, that the lessons learned from an exactly solvable system
will help to shed additional light on
the debate, and that they will aid in building much needed intuition about the appropriate mix
of quantum and classical in our understanding of the dynamics that play a role in a plethora of
physical contexts.

This material is based upon work supported by the Air Force Office of Scientific Research under
award number FA9550-18-1-0183.



%

\end{document}